\documentclass[%
 aip,
 amsmath,amssymb,
 reprint,%
]{revtex4-1}

\usepackage{graphicx}
\usepackage{dcolumn}
\usepackage{bm}

\usepackage[utf8]{inputenc}
\usepackage[T1]{fontenc}
\usepackage{mathptmx}
\usepackage{etoolbox}
\usepackage{graphicx}
\usepackage{dcolumn}
\usepackage{bm}
\usepackage{enumitem}
\usepackage{amssymb}
\usepackage{amsmath}
\usepackage{psfrag}
\usepackage{mathtools}
\usepackage{color}
\usepackage{xcolor}
\usepackage[export]{adjustbox}
\usepackage[english]{babel}

\usepackage{natbib}

\definecolor{cyan1}{rgb}{0.387, 0.82, 1}
\definecolor{red1}{rgb}{0.902, 0.383, 0.355}
\definecolor{RGBred}{rgb}{1,0,0}
\definecolor{RGBblue}{rgb}{0,0,1}
\definecolor{outgreen}{rgb}{0.38,0.73,0.66}
\definecolor{arrowgreen}{rgb}{0,0.631,0.294}
\definecolor{arrowblue}{rgb}{0.129,0.251,0.6}
\definecolor{arrowgreen}{rgb}{0,0.631,0.294}
\definecolor{arrowred}{rgb}{0.929,0.11,0.141}
\definecolor{poleorange}{rgb}{0.984,0.69,0.231}
\definecolor{poleblue}{rgb}{0.161,0.667,0.886}
\definecolor{planegray}{rgb}{0.502,0.502,0.502}

\newcommand{\tr}[1]{{#1}} 
\newcommand{\sr}[1]{{#1}} 
\newcommand{\fc}[1]{{#1}}
\newcommand{\dc}[1]{{#1}}
\newcommand{\lc}[1]{{#1}}
\newcommand{\cc}[1]{{#1}}

\renewcommand{\arraystretch}{1.3}

\makeatletter
\def\@email#1#2{%
 \endgroup
 \patchcmd{\titleblock@produce}
  {\frontmatter@RRAPformat}
  {\frontmatter@RRAPformat{\produce@RRAP{*#1\href{mailto:#2}{#2}}}\frontmatter@RRAPformat}
  {}{}
}%
\makeatother
\begin{document}

\preprint{AIP/123-QED}

\title{\sr{Exact potentials in multivariate Langevin equations}}%

\author{Tiemo Pedergnana}
 \email{ptiemo@ethz.ch}
 \affiliation{%
 CAPS Laboratory, Department of Mechanical and Process Engineering, ETH Z{\"u}rich, Sonneggstrasse 3,
8092 Z{\"u}rich, Switzerland
}%
\author{Nicolas Noiray}%
 \email{noirayn@ethz.ch}
\affiliation{%
 CAPS Laboratory, Department of Mechanical and Process Engineering, ETH Z{\"u}rich, Sonneggstrasse 3,
8092 Z{\"u}rich, Switzerland
}%

\date{\today}

\begin{abstract}
\sr{Systems governed by a multivariate Langevin equation featuring an exact potential exhibit straightforward dynamics but are often difficult} to recognize because, after a general coordinate \tr{change}, the gradient flow becomes obscured by the Jacobian matrix of the mapping. \dc{In this work, a detailed analysis of the transformation properties of Langevin equations under general nonlinear mappings is presented.} We show how to identify systems \dc{with exact potentials} \tr{by} understanding their \dc{differential-geometric} properties. \dc{To demonstrate the power of our method, we use it to} derive exact potentials for broadly studied \tr{models} of nonlinear deterministic and stochastic oscillations. \dc{In selected examples, we visualize the identified potentials. Our results imply a broad class of exactly solvable stochastic models which can be self-consistently defined from given deterministic gradient systems.}
\end{abstract}

\maketitle

\dc{\begin{quotation}
Since the seminal works of Albert Einstein and Paul Langevin on Brownian motion, the study of stochastic dynamics has developed into a fruitful science which today finds application in many fields. For an important subclass of noise-driven systems, those governed by a Langevin equation with an exact stationary potential, the steady-state dynamics may be solved analytically. Despite the great benefits of these solutions, the subtleties of identifying exact potentials in systems which are described in transformed variables have apparently been ignored so far. To fill this gap in the literature, we derive the differential-geometric transformation properties of multivariate Langevin equations under general coordinate changes and we demonstrate how they can lead to new analytical descriptions of a system's nonlinear dynamics. The method is then applied to different examples of deterministic and noise-driven oscillations. Finally, we comment on a broad class of exactly solvable models implied by our results, which enable self-consistent and analytical modeling of additive \cc{white} noise in given deterministic gradient flows.
\end{quotation}}

\section{Introduction \label{Introduction}}
\sr{In systems \fc{featuring} exact potentials,} the evolution of a $n$-dimensional set of variables ${x}=(x_1,\dots,x_n)^T$: $\mathbb{R}\rightarrow \mathbb{R}^n$ over time $t\in\mathbb{R}$ \tr{is} governed by the Langevin equation \sr{with potential} (\sr{LP})\cite{gardiner1985handbook,risken1996fokker}
\begin{eqnarray}
    \dot{{x}}=-\nabla \mathcal{V}({x},t)+{\Xi}, \label{Langevin equation}
\end{eqnarray}
where \cc{$\dot{(\,\,)}=d(\,\,)/dt$ is the total derivative with respect to time,} \cc{$\nabla=(\partial/\partial x_1,\dots,\partial/\partial x_n)^T$ is the gradient operator}, $\mathcal{V}$: $\mathbb{R}^n\cc{\times \mathbb{R}}\rightarrow \mathbb{R}$ is the potential, \sr{\cc{$\mathcal{F}_i=-\partial\mathcal{V}(x,t)/\partial x_i$} is the $i$th component of the restoring force $\mathcal{F}$} and the vector ${\Xi}=(\xi_1,\dots,\xi_n)^T$: $\mathbb{R}\rightarrow \mathbb{R}^n$ contains white \cc{Gaussian} noise sources $\xi_i$, $i=1,\dots,n$ of equal intensity \cc{(variance)} $\Gamma$ \cc{and zero mean}.\cc{\footnote{\cc{In the following, we assume all white noise sources considered here to be Gaussian with zero mean.}}} The individual entries $\xi_i$ of $\Xi$ are assumed to be delta-correlated: $\langle \xi_i \xi_{i,\tau} \rangle=\Gamma \delta{(\tau)}$, where $\langle\cdot \rangle$ is the expected value operator, $(\cdot)_{,\tau}$ denotes a positive time shift by $\tau$ and $\delta$ is the Dirac delta function.\cite{stratonovich1963topics} 

The modern study of Langevin equations dates back over a century,\cite{Hanggi2005} and \cc{continues} to be an active topic of research today.\lc{\cite{Zaks2005,Nakao2010,Yamapi2012,Frishman2020,SantosGutierrez2021}} \sr{Well known, low-dimensional examples of LPs are the Stuart--Landau oscillator\cite{Stuart19581,landau1959fluid,Lax1967290} \lc{subject to additive \cc{white} noise, whose deterministic part} represents the normal form of a supercritical Hopf bifurcation,\cite[see][p. 270]{arnold2012geometrical} and the deterministically and stochastically averaged noise-driven Van der Pol oscillator.\cite{van1926lxxxviii,sanders2007averaging,stratonovich1963topics,Roberts1986111,balanov2009simple} \lc{As will be discussed in the present work, m}ultivariate systems governed by potentials are also found} in \sr{the classic} Kuramoto model,\cite{kuramoto1984chemical,Strogatz20001} swarming oscillators,\cite{okeefe} networks of coupled limit cycles\cite{Aronson1990403,balanov2009simple,Pedergnana20221133} and in models of noise-driven, self-sustained \lc{thermoacoustic} modes of annular cavities.\cite{Noiray2013,Faure-Beaulieu2020,Ghirardo2021,indlekofer_faure-beaulieu_dawson_noiray_2022} \lc{We mention that exact potentials occur also in models of turbulent wakes,\cite{Rigas2015} swirling flows\cite{Sieber2021} and buoyancy-driven bodies.\cite{Tchoufag2015}}

\dc{This work deals with noise-driven systems in the form of ordinary \lc{stochastic} differential equations. The reader interested in applications of exact potentials in partial differential equations is referred to the relevant literature.\cite{Graham19904661,Graham1991,Descalzi1994509,Montagne199647,Izus199893}}

\sr{In general, multivariate dynamical systems do not possess an exact potential. While the problem of finding meaningful quasi-potentials in systems that are \textit{not} governed by an exact potential has been tackled in the past,\dc{\cite{Zhou20123539,Cameron20121532}} there are also many relevant multivariate systems subject to random noise \textit{with} exact potentials. If one exists, \cc{and it is stationary}, then knowing the exact potential is a great benefit because it fully determines the stochastic dynamics in the steady state.\cc{\footnote{\cc{While there is undoubtedly also value in knowing the exact \textit{unsteady} potential, if present, in a multivariate Langevin equation with explicitly time-dependent deterministic part $\mathcal{F}$, a further investigation of such systems and the role of their potentials is left for future research.}}} The problem practitioners face is that it is often difficult to perceive the existence of a potential when the system is described \cc{in} transformed variables. This issue is addressed in this work.}

\begin{figure}[t!]
\begin{psfrags}
    
\psfrag{a}{$x=f(y)$}
\psfrag{b}{\hspace{-0.1cm}$\mathcal{D}\hspace{3.7cm}f^{-1}(\mathcal{D})$}
\psfrag{c}{\hspace{-0.1cm}$0.2$\hspace{1.8cm}$0.6$\hspace{1.8cm}$5.4$}
\psfrag{d}{\hspace{-0.25cm}$0$\hspace{2.4cm}$4$\hspace{2.4cm}$8$}
\psfrag{e}{\hspace{1cm}detuning $\Delta/\lambda$ (-)}
\psfrag{1}{$A$}
\psfrag{2}{\hspace{0.08cm}$\Phi$}
\psfrag{3}{$t$}

    \centering
    \hspace{-0.05cm}\includegraphics[width=0.3\textwidth]{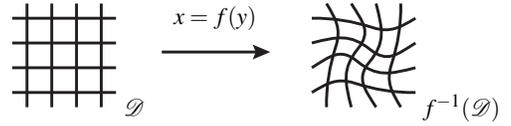}
\end{psfrags}
    \caption{\lc{Cartesian coordinates $x\in\mathcal{D}$ and their transformed counterparts $y \in f^{-1}(\mathcal{D})$, both represented in the $x$-frame.} If potential systems driven by additive \cc{white} noise are described in transformed variables, the corresponding exact potentials can be identified through the systems' differential-geometric properties, which are studied in this work.}
    \label{Figure 1}
\end{figure}

In particular, we are concerned with identifying the \sr{presence of an underlying exact potential} in general noise-driven systems taking the form 
\begin{eqnarray}
    \dot{{x}}=\mathcal{F}({x},t)+\mathcal{B}({x}){\Xi}, \label{General noise-driven system}
\end{eqnarray}
where $\mathcal{F}$ is a vector- and $\mathcal{B}$ a $n$-by-$n$ tensor field.\cite{gardiner1985handbook} With the knowledge of $\mathcal{F}$, assuming a LP \eqref{Langevin equation}, one can easily deduce if an exact potential $\mathcal{V}$ exists for $x$ by checking the following necessary and sufficient conditions\dc{\cite[see][pp. 133--134]{risken1996fokker}}:
\sr{\begin{eqnarray}
    \nabla_i \mathcal{F}_j=\nabla_j \mathcal{F}_i \label{symmetry condition}
\end{eqnarray}
for all $i$ and $j\neq i$. However, if \dc{these conditions are} not fulfilled, this does not preclude the existence of an exact potential governing the original variables that were \cc{\textit{transformed}} into $x$ via a certain nonlinear mapping. We therefore argue that, instead of applying Eq. \eqref{symmetry condition},} Eq. \eqref{General noise-driven system} should be compared to a \sr{LP} \tr{after a coordinate change defined by an} arbitrary nonlinear mapping 
\begin{eqnarray}
{x}={f}({y}), \label{nonlinear transformation}
\end{eqnarray}
\dc{see Fig. \ref{Figure 1}.} Assuming \sr{purely additive \cc{white} noise in the equations governing the underlying potential system which transforms objectively under local rotations and reflections,} the resulting transformed Langevin equation \sr{with potential} (\sr{TLP}) reads, after redefining $y\rightarrow x$,
\begin{eqnarray}
    \dot{x}=-{g}^{-1}(x)\nabla \widetilde{\mathcal{V}}(x,t)+{h}^{-1}(x)\Xi, \label{Transformed Langevin equation}
\end{eqnarray}
where the Jacobian of $f$, 
\begin{eqnarray}
    J(x)=\nabla {f(x)},
\end{eqnarray}
was assumed to be nonsingular (invertible) with polar decomposition\cc{\cite[see][p. 449]{horn2012matrix}} \sr{$J=Qh$}, $Q=Q^{-T}$ is orthogonal, \cc{$h$ is a positive definite matrix,} $g=h^T h$ is the \cc{symmetric,} positive definite metric tensor\dc{, $(\cdot)^T$ is the transpose} and 
\begin{eqnarray}
    \widetilde{\mathcal{V}}(x,t)=\mathcal{V}\big(f(x),t\big)
\end{eqnarray} 
is the transformed potential.\dc{\footnote{Unless explicitly stated otherwise, all quantities in this work are assumed to be real.}}

In this work, we derive necessary and sufficient conditions for the existence of an exact potential in a noise-driven system given by Eq. \eqref{General noise-driven system}.\cc{\footnote{\cc{Note that the concept of studying the transformation properties of Eq. \eqref{Langevin equation} for \textit{specific} systems is not novel. In fact, Eq. \eqref{Transformed Langevin equation} generalizes a result which was stated in our earlier work without an explicit derivation.\cite{Pedergnana20221133} The present work is the first instance, however, where this method is proposed for identifying exact potentials in \textit{general} noise-driven systems given by Eq. \eqref{General noise-driven system}.}}} After briefly recalling some of the special properties of \sr{potential systems driven by purely additive \cc{white} noise (PANs),} we study their transformation rules under the mapping $f$ from a continuum-mechanical perspective before applying our results to broadly studied examples of nonlinear oscillatory systems.

\section{Noise-driven potential flows}
\dc{\subsection{Special properties}}
In this section, we list a few of the simplifications, compared to general dynamical systems, which offer themselves for PANs. We first analyze the stochastic case with \sr{$\Gamma\neq 0$} using the Fokker-Planck equation (FPE), which describes the evolution of the joint probability density function \cc{(PDF)} $P$: $\mathbb{R}^n\cc{\times \mathbb{R}}\rightarrow \mathbb{R}$ of a random dynamic variable $x$ over time.\cite{gardiner1985handbook,risken1996fokker} The FPE associated with the \sr{LP} \eqref{Langevin equation} reads 
\begin{eqnarray}
    \frac{\partial P({x},t)}{\partial t}=\nabla \cdot \big[P({x},t)\nabla \mathcal{V}({x},t)+ \frac{\Gamma}{2}\nabla P({x},t)\big]. \label{FP equation}
\end{eqnarray}
\noindent \lc{Assuming a steady potential $\partial \mathcal{V}/\partial t=0$}, we make the substitution
\begin{eqnarray}
    P({x},t)=G({x},t) \exp{\bigg(\dfrac{-2 \mathcal{V}({x})}{\Gamma}\bigg)}, \label{Ansatz}
\end{eqnarray}
which leads to an advection--diffusion equation for $G$:
\begin{eqnarray}
     \frac{\partial G({x},t)}{\partial t}+{v}(x)\cdot \nabla G({x},t)=\dfrac{\Gamma}{2}\nabla^2 G({x},t), \label{advection diffusion equation}
\end{eqnarray}
\lc{where $v(x)=-\nabla \mathcal{V}(x)$ is the velocity field of the gradient flow.} \tr{Equation \eqref{advection diffusion equation} is analytically solvable in special cases.\cite{Philip19943545,Zoppou1997144,Zoppou1999667}} Of special interest is the exact solution of Eq. \eqref{advection diffusion equation} given by $G=\text{const.}$, which corresponds to the \cc{steady-state} PDF $P({x},t\rightarrow \infty)=P_{\infty}({x})$:
\begin{eqnarray}
    P_\infty({x})=\mathcal{N} \exp{\bigg(\dfrac{-2 \mathcal{V}({x})}{\Gamma}\bigg)}, \label{Steady FP solution}
\end{eqnarray}
where $\mathcal{N}\in\mathbb{R}^+$ is a normalization constant. As shown below, the \cc{transformed} \cc{steady-state} PDF \cc{$\widetilde{P}_\infty(x)=P_\infty\big(f(x)\big)$} of the \sr{TLP} \tr{\eqref{Transformed Langevin equation}} can be derived in analogous fashion:
\begin{eqnarray}
    \tr{\widetilde{P}_\infty(x)=\mathcal{N}\exp{\bigg(\dfrac{-2 \widetilde{\mathcal{V}}({x})}{\Gamma}\bigg)}.} \label{transformed steady sol}
\end{eqnarray}
In the deterministic limit $\Gamma=0$ \sr{and for \cc{a general,} time-varying potential}, the \sr{TLP} \tr{\eqref{Transformed Langevin equation}} is reduced to the transformed gradient system \dc{(TGS)}
\begin{equation}
    \dot{x}=-g^{-1}(x)\nabla \widetilde{\mathcal{V}}(x,t). \label{gradient flow}
\end{equation}
Equation \eqref{gradient flow} states that trajectories $x(t)$ are attracted to lower values of the potential $\widetilde{\mathcal{V}}$, the attraction being \tr{equal} to the potential gradient scaled with the inverse metric tensor $g^{-1}$\cc{, a positive definite matrix}. By definition and by the positive definiteness of $g$, \lc{any stationary potential} $\widetilde{\mathcal{V}}$\lc{, $\partial \widetilde{\mathcal{V}}/\partial t$=0,} is a Lyapunov function of the variables $x$ evolving under Eq. \eqref{gradient flow} and thus determines the local and global stability of its trajectories.\cite{Lyapunov1992531}
\dc{\subsection{Self-consistent modeling}}
\dc{Given a system of the form 
\begin{eqnarray}
\dot{x}=-\mathcal{M}(x)\nabla \widetilde{\mathcal{V}}(x,t),
\end{eqnarray}
where $\mathcal{M}$ is an arbitrary positive definite matrix, one can directly identify a TGS by defining $g^{-1}=\mathcal{M}$. Furthermore, we recall that the Cholesky decomposition of a real positive definite matrix is, for each $x$, uniquely defined as
\begin{eqnarray}
\mathcal{M}(x)=\mathcal{L}(x)\mathcal{L}^T(x), \label{L decomp}
\end{eqnarray}
where $\mathcal{L}$ is a lower triangular matrix with positive diagonal entries.\cite[see][p. 441]{horn2012matrix} \dc{Since the diagonal of a triangular matrix contains its eigenvalues, $\mathcal{L}$ is also positive definite.} Comparing Eq. \eqref{L decomp} to the definition 
\begin{eqnarray}
    g^{-1}(x)=h^{-1}(x)h^{-T}(x)
\end{eqnarray}
and setting $h^{-1}=\mathcal{L}$, one can define the following corresponding PAN:
\begin{eqnarray}
    \dot{x}=-\mathcal{M}(x)\nabla \widetilde{\mathcal{V}}(x,t)+\mathcal{L}(x) \Xi,
\end{eqnarray}
where $\Xi$ is a vector containing white noise sources of equal intensity. For a stationary potential $\partial \mathcal{V}/\partial t=0$, under the above assumptions, the exact \cc{(transformed)} steady-state PDF of this system is given by Eq. \eqref{transformed steady sol}.}
\section{Transformation rules}
\subsection{Gradient flow}
We now derive the transformation rules for \sr{PANs given by Eq. \eqref{Langevin equation}}, beginning with the gradient term. We use the Einstein summation convention\dc{,} by which repeated indices in a product imply summation over these indices. In index form, the \sr{LP} \eqref{Langevin equation} with $\Xi=0$ reads 
\begin{eqnarray}
    \dot{x}_k=-\frac{d \mathcal{V}(x,t)}{d x_k}, \label{Gradient system}
\end{eqnarray}
for $k=1,\dots,n$, where the partial $x$-derivatives in $\nabla$ have been rewritten as total derivatives because $\mathcal{V}$ depends only on the (spatially) independent variables \tr{$x$} and $t$. Under the the transformation $x=f(y)$, suppressing for brevity the dependence of $f$ on $y$ in the argument of $\mathcal{V}$, Eq. \eqref{Gradient system} becomes
\begin{eqnarray}
    \frac{d f_k(y)}{d y_i} \frac{d y_i}{ d t}&=&-\frac{d y_j}{d x_k}\frac{d \mathcal{V}(f,t)}{d y_j} \nonumber \\
    &=&-\frac{d y_j}{d f_k (y)}\frac{d\mathcal{V}(f,t)}{d y_j}, \label{Jacobian xdot}
\end{eqnarray}
which can be rewritten as
\begin{eqnarray}
    \frac{d y_i}{d t}
    &=&-\frac{d y_i}{d f_k(y)}\frac{d y_j}{d f_k(y)}\frac{d\mathcal{V}(f,t)}{d y_j}.\label{Intermediate result}
\end{eqnarray}
The formula for the squared length of an infinitesimal line element $d s^2$ in general curvilinear coordinates $y$ is 
\begin{eqnarray}
    d s^2=g_{i j}d y_i d y_j, \label{g in general coordinates}
\end{eqnarray}
 and the value of this quantity is independent of the coordinate system.\cc{\cite[see][pp. 213--214]{kuhnel2015differential}} \sr{To relate $ds^2$ to the original coordinates $x$, we consider the case where the mapping $f$ is simply the identity: $x=y$. This gives $g_{i j}=I_{i j}$, where $I$ is the identity matrix, so that 
 \begin{eqnarray}
     d s^2=d x_k d x_k. \label{g in Cartesian coordinates}
 \end{eqnarray}}
Multiplying both sides of Eq. \eqref{Intermediate result} with $g_{ij}$\sr{, using \cc{Eqs. \eqref{g in general coordinates}--\eqref{g in Cartesian coordinates}}} and noting that $df_k =dx_k $, we obtain
\begin{eqnarray}
    g_{ij}\frac{d y_i}{d t}
    &=&-\frac{d \mathcal{V}(f,t)}{d y_j}, \label{int. res. 1}
\end{eqnarray}
which can, by the symmetry of the metric tensor $g$, be written in vector form as follows:
\begin{eqnarray}
   \dot{y}
    &=&-g^{-1}(y) \nabla_y  \widetilde{\mathcal{V}}(y,t), \label{int. res. 2}
\end{eqnarray}
where $(\nabla_y)_i=\partial/\partial y_i$ and the transformed potential was defined as $\widetilde{\mathcal{V}}(y,t)=\mathcal{V}\big(f(y),t\big)$. As before, we interchanged partial and total derivatives in going from Eq. \eqref{int. res. 1} to Eq. \eqref{int. res. 2} because $\widetilde{\mathcal{V}}$ depends on $y$ solely through $f$, and therefore the chain rule is the same for $\partial \widetilde{\mathcal{V}}/\partial y_j$ as for $d \widetilde{\mathcal{V}}/d y_j$, i.e., the two terms coincide:
\begin{eqnarray}
    \frac{d \widetilde{\mathcal{V}}(f,t)}{d y_j}&=&\frac{d \widetilde{\mathcal{V}}(f,t)}{d f}\frac{d f(y)}{d y_j}\\
    &=&\frac{\partial \widetilde{\mathcal{V}}(f,t)}{\partial f}\frac{\partial f(y)}{\partial y_j}.
\end{eqnarray}
We then infer from Eq. \eqref{int. res. 2} that under the mapping $x=f(y)$, the potential gradient transforms like
\begin{eqnarray}
    -\nabla \mathcal{V}(x,t)\rightarrow-g^{-1}(y)\nabla_y \widetilde{\mathcal{V}}(y,t). \label{gradient flow trafo formula}
\end{eqnarray}
Redefining $y\rightarrow x$ and $\nabla_y\rightarrow \nabla$ in Eq. \eqref{int. res. 2} yields Eq. \eqref{gradient flow}.

\subsection{Noise term \label{noise term section}}
Having studied the transformation properties of the \sr{deterministic} gradient flow, we now turn to the noise term $\Xi$ in Eq. \eqref{Langevin equation}. Knowing from Eq. \eqref{Jacobian xdot} that under the mapping $x=f(y)$, $\dot{x}_k=J_{k j} \dot{y}_j$, where $J_{k j}=\partial f_k/\partial y_j$, we can infer that $\Xi$ transforms like
\begin{eqnarray}
   \sr{\Xi \rightarrow J^{-1}(y) \Xi\big(\widetilde{\Xi}\big).} \label{general trafo}
\end{eqnarray}
The problem one now faces is that it is not clear a priori how $\Xi$ is related to the transformed noise vector $\widetilde{\Xi}$\cc{,} i.e., the noise vector in $y$-coordinates. To resolve this issue, we assume that \sr{$\widetilde{\Xi}$} preserves the noise intensity and the local orientation \sr{of \cc{the additive white noise} $\Xi$ in the original coordinates}. In other words, \cc{under $x=f(y)$}, $\Xi$ behaves like an objective vector field transformed by an orthogonal tensor field $Q=Q^{-T}$ representing the local rotation \sr{or reflection} associated with $f$:\cc{\cite[see][p. 42]{truesdell2004non}}
\begin{eqnarray}
   \sr{\Xi= Q(y) \widetilde{\Xi}.} \label{objective trafo}
\end{eqnarray}
Note that $Q$ can be directly obtained from the polar decomposition of the Jacobian of $f$:
\begin{eqnarray}
    J(y)=Q(y)h(y), \label{Polar decomposition}
\end{eqnarray}
where $h$ is a positive definite matrix (since $J$ is, by assumption, invertible) of the same size as $J$ and $Q$. \sr{Using Eqs. \eqref{objective trafo}--\eqref{Polar decomposition}, the transformation formula \eqref{general trafo} can be simplified as follows:}
\begin{eqnarray}
    \Xi \rightarrow h^{-1}(y) \widetilde{\Xi}. \label{noise trafo formula}
\end{eqnarray}
Combining \eqref{gradient flow trafo formula} and \eqref{noise trafo formula} yields the \sr{TLP} for systems with purely additive \cc{white} noise \cc{satisfying Eq. \eqref{objective trafo}}:
\begin{eqnarray}
    \tr{\dot{y}=-{g}^{-1}(y)\nabla_y\widetilde{\mathcal{V}}(y,t)+{h}^{-1}(y)\widetilde{\Xi}.} \label{Transformed Langevin equation, force-like noise}
\end{eqnarray}
Redefining $y\rightarrow x$, $\nabla_y\rightarrow \nabla$ and $\widetilde{\Xi}\rightarrow \Xi$ reduces Eq. \eqref{Transformed Langevin equation, force-like noise} to Eq. \eqref{Transformed Langevin equation}. \cc{Note that, throughout this work, although $\Xi$ appears in the TLP \eqref{Transformed Langevin equation} as a multiplicative noise, we nevertheless refer to it as additive because it is clear from the above discussion that it derives from a purely additive noise term in the LP \eqref{Langevin equation} formulated in the original coordinates, and that its multiplicative character is solely due the system's representation in transformed variables.}

We stress that the assumptions made on the noise term in order to obtain Eq. \eqref{objective trafo} are not trivial, and that there may be situations where \cc{noise terms} appear that do not transform according to the same formula. \cc{As an example for such a term, consider the case where $\Xi$ is given by an Ornstein--Uhlenbeck process \cite{Uhlenbeck1930823} satisfying 
\begin{eqnarray}
    \dot{\Xi}=-\dfrac{\Xi}{\vartheta}+\zeta, \label{O-U noise}
\end{eqnarray}
where $\vartheta$ is the correlation time and $\zeta$ is an additive white noise as defined in Sec. \ref{Introduction}. If $\zeta$ transforms according to Eq. \eqref{objective trafo}, because of the time-differentiation in Eq. \eqref{O-U noise}, the same will in general not be true for $\Xi$.}\cc{\footnote{\cc{Alternatively, if $\Xi$ is governed by Eq. \eqref{O-U noise} and \textit{does} transform according to Eq. \eqref{objective trafo}, then the resulting transformed noise $\widetilde{\Xi}$ is not an Ornstein--Uhlenbeck process in general.}}} Nevertheless, the special case described by Eq. \eqref{Transformed Langevin equation} correctly identifies the exact potentials in the examples presented below\cc{, which exclusively feature white noise}. 

\subsection{Fokker--Planck equation}
The probability $\mathcal{P}$ of the state $x$ being inside the domain $\mathcal{D}$ at time $t$ is defined as
\begin{eqnarray}
    \mathcal{P}=\int_\mathcal{D} P(x,t) dV,\label{Probability}
\end{eqnarray}
where $dV=dx_1,\dots,dx_n$ is the volume element. In ${y}$-coordinates, using $\mathrm{det}(J)=\mathrm{det}(h)$, where $\mathrm{det}$ is the determinant, Eq. \eqref{Probability} can be rewritten as 
\begin{eqnarray}
    \mathcal{P}=\int_{\widetilde{\mathcal{D}}} P\big({f}({y}),t\big) \big|\mathrm{det}\big(h(y)\big)\big| d\widetilde{V}, \label{transformed volume integral}
\end{eqnarray}
where $d\widetilde{V}=dy_1,\dots,dy_n$ is the transformed volume element and $\widetilde{\mathcal{D}}=f^{-1}(\mathcal{D})$ is the transformed domain. We learn from Eq. \eqref{transformed volume integral} that the PDF $P$ transforms like
\begin{eqnarray}
   \tr{{P}(x,t)}\rightarrow\big|\mathrm{det}\big(h(y)\big)\big|{P}\big(f(y),t\big) \label{Transformed stationary FP solution}
\end{eqnarray}
under the mapping $x=f(y)$. Given a \sr{TLP} \eqref{Transformed Langevin equation}, one can directly obtain $\mathrm{det}(h)$ by computing the determinant of the matrix ${h}^{-1}$ and using $\mathrm{det}(h^{-1})=\mathrm{det}(h)^{-1}$. Typically, however, there is no interest in this geometric prefactor and the quantity of importance is the transformed PDF 
\begin{eqnarray}
    \widetilde{P}(y,t)=P\big(f(y),t\big).
\end{eqnarray}
Comparing Eq. \eqref{Steady FP solution} to Eq. \eqref{Transformed stationary FP solution}, we observe that the steady-state solution $P_\infty$ of the \cc{FPE \eqref{FP equation} associated with the} \sr{LP} \eqref{Langevin equation} with stationary potential transforms under $f$ like
\begin{eqnarray}
    P_\infty({x})\rightarrow \big|\mathrm{det}\big(h(y)\big)\big|\widetilde{P}_\infty(y). \label{Steady FP solution trafo}
\end{eqnarray}
Knowledge of $\widetilde{\mathcal{V}}$ is sufficient to deduce the transformed \cc{steady-state} PDF
\begin{eqnarray}
    \widetilde{P}_\infty(y)=\mathcal{N}\exp{\bigg(\dfrac{-2 \widetilde{\mathcal{V}}({y})}{\Gamma}\bigg)},
\end{eqnarray}
which, after redefining $y\rightarrow x$, coincides with Eq. \eqref{transformed steady sol}.

\section{Potential identification \label{method section}}
In the stochastic case, if $\Xi$ is an \sr{additive \cc{white} noise vector with nonzero entries satisfying the assumptions made in the previous section} and $\mathcal{B}$ is nonsingular, identifying the exact potential $\widetilde{\mathcal{V}}$ \cc{in} a general noise-driven system \tr{\eqref{General noise-driven system}} is straightforward. By comparison with the \sr{TLP} \eqref{Transformed Langevin equation}, $h$ and $g=h^Th$ are directly obtained from Eq. \eqref{General noise-driven system}: 
\begin{eqnarray}
    h(x)&=&\mathcal{B}^{-1}(x),\\
    g(x)&=&\mathcal{B}^{-T}(x)\mathcal{B}^{-1}(x).
\end{eqnarray}
We recall that three-dimensional \sr{potential systems} are uniquely defined by the vector identity \begin{eqnarray}
    \mathrm{curl}\,\,\mathrm{grad} (\cdot)=0,
\end{eqnarray}  
i.e., the curl of a vector field is zero if and only if the vector field can be written as the gradient of a scalar function. Generalized to arbitrary dimensions, the equivalent identity reads 
\begin{eqnarray}
    \mathrm{skew}\big[\cc{H} (\cdot)\big]=0,
\end{eqnarray} 
where $\mathrm{skew}(\cdot)$ is the skew-symmetric part and $\cc{H}(\cdot)$ is the Hessian matrix. By comparing Eqs. \eqref{General noise-driven system} and \eqref{Transformed Langevin equation} and using \cc{$g=h^T h$}, then, the following criteria are readily deduced:
\begin{itemize}
    \item[\tr{(I)}] A general noise-driven system described by Eq. \eqref{General noise-driven system} has an exact potential if and only if there exists a positive definite symmetric tensor field $\mathcal{M}$ such that $$\mathrm{skew}({\nabla[ \mathcal{M}\dc{^{-1}}(x) \mathcal{F}(x)]^{\cc{T}}})$$
    \cc{is equal to a $n$-by-$n$ zero matrix.}
     \item[\tr{(II)}] For \sr{nonzero additive} noise $\Xi\neq 0$ \sr{satisfying the assumptions made in Sec. \ref{noise term section}} and nonsingular $\mathcal{B}$, a noise-driven system given by Eq. \eqref{General noise-driven system} has an exact potential if and only if $$\mathrm{skew}({\nabla[ \mathcal{B}^{-T}(x) \mathcal{B}^{-1}(x) \mathcal{F}(x)]^{\cc{T}}})$$
     \cc{is equal to a $n$-by-$n$ zero matrix.}
\end{itemize}
If either \tr{(I)} or \tr{(II)} are satisfied, the term in the square bracket is proportional to the \cc{(negative)} potential gradient. If \tr{(I)} is satisfied, $\mathcal{M}$ is proportional to the \dc{inverse} metric tensor \dc{$g^{-1}$.} Note that in the purely deterministic case \cc{with} $\Xi=0$, \tr{(II)} does not apply \dc{because in this case, $\mathcal{B}$ is ill-defined}. 

The general criterion \tr{(I)} involves the solution of an underdetermined system of \cc{partial} differential equations for a matrix $\mathcal{M}$ whose entries are constrained by its symmetry and positive definiteness, and is impractical for manual analysis. In the future, this criterion may be simplified or solved with computer algebra. \cc{Currently, in} practice, \dc{finite}-dimensional gradient systems with $\Xi=0$ are identified by inspection of Eq. \eqref{General noise-driven system} under consideration of the \tr{general} form of a \sr{TLP} \eqref{Transformed Langevin equation}. Specific examples are discussed below.

\begin{figure}[t!]
\begin{psfrags}
    
\psfrag{a}{}
\psfrag{b}{\hspace{-1.2cm}forcing amplitude \lc{$\log_{10}(F/\omega\lambda^{3/2})$}}
\psfrag{c}{\lc{\hspace{-0.4cm}$-0.903$\hspace{1.51cm}$0$\hspace{1.51cm}$0.903$}}
\psfrag{d}{\lc{\hspace{-0.25cm}$0$\hspace{1.97cm}$2$\hspace{1.97cm}$4$}}
\psfrag{e}{\hspace{0.9cm}detuning $\Delta/\lambda$ }
\psfrag{1}{$A$}
\psfrag{2}{\hspace{0.045cm}$\Phi$}
\psfrag{3}{$t$}

    \centering
    \hspace{-0.05cm}\includegraphics[width=0.43\textwidth]{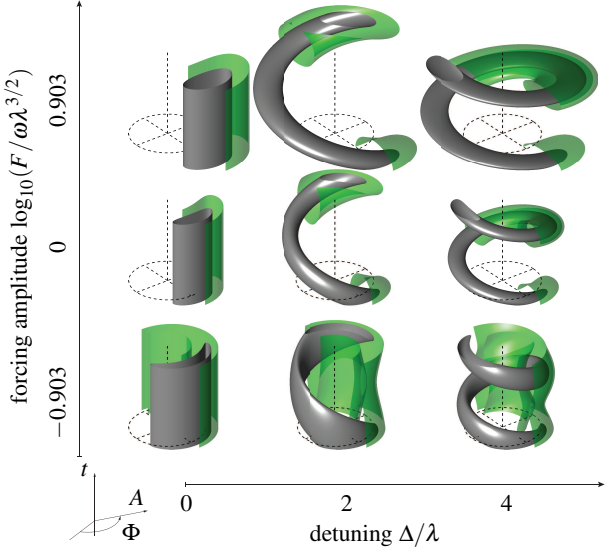}
\end{psfrags}
    \caption{\dc{Illustration of the results derived in Sec. \ref{vdp section}. Shown are isosurfaces of the time-dependent \lc{potential} $\widetilde{\mathcal{V}}$ defined by Eq. \eqref{vdp potential} corresponding to $60\%$ (green, \lc{one half shown}) and $90\%$ (gray) of its \cc{(negative)} \lc{minimum} value, as a function of the nondimensionalized forcing amplitude \lc{$F/\omega\lambda^{3/2}$} and detuning $\Delta/\lambda$\cc{, respectively} (semi-$\log$ scale). Parameter values are given in the main text. The length of the dashed vertical line is $2/\lambda$ and the dashed circle's radius is equal to the \cc{unforced} limit cycle amplitude $2\sqrt{\lambda}$.}}
    \label{Figure 2}
\end{figure}

\section{Examples \label{Example section}}
We now demonstrate our method on some of the examples mentioned in Sec. \ref{Introduction}. \dc{In certain cases, we visualize the identified potentials for different parameter values. A more in-depth analysis of these systems' nonlinear dynamics, which is out of the scope of this work, is left for future research.} All equations are presented in the same form in which they appear in the references, up to minor changes in notation.
\subsection{\sr{Averaged Van der Pol oscillator} \label{vdp section}}
The \sr{weakly nonlinear dynamics of a harmonically forced, noise-driven \sr{Van der Pol} oscillator \sr{synchronized with} the forcing frequency $\omega$ can be derived using deterministic and stochastic averaging.} \dc{The resulting equations are\cite[see][Eqs. (7.58) and (7.59)\lc{\footnote{\lc{Due to a typographical error, the ``$A^2$'' in the bracket is missing in Eq. (7.58) of the reference.}}}]{balanov2009simple}:}
\begin{eqnarray}
    \dot{A}&=&\frac{A}{2}(\lambda-\frac{A^2}{4})-\frac{F}{2\omega}\sin\varphi +\frac{\Gamma}{4 \omega^2 A}+\eta_1,\label{example 1 eq 1}\\
    \dot{\varphi}&=&\Delta-\frac{F}{2\omega A}\cos\varphi+\frac{\eta_2}{A}, \label{example 1 eq 2}
\end{eqnarray}
where $\Delta=(\omega_0^2-\omega^2)/2\omega\approx \omega_0-\omega$ is the detuning between the eigen- ($\omega_0$) and  the forcing ($\omega$) frequency, $2\sqrt{\lambda}$ is the \cc{unforced} limit cycle amplitude, $F$ is the forcing amplitude, \tr{$\Xi=(\eta_1,\eta_2)^T$} and $\eta_{1,2}$ are white noise sources of equal intensity $\Gamma/2\omega^2$.

\cc{To identify the transformed potential $\widetilde{\mathcal{V}}$, we define} the new variables $x=(A,\Phi)^T$, where $\Phi=\varphi-\Delta t$\cc{, and rewrite Eqs. \eqref{example 1 eq 1}--\eqref{example 1 eq 2} as
\begin{eqnarray}
    \dot{x}&=&\underbrace{\begin{pmatrix}
\frac{A}{2}(\lambda-\frac{A^2}{4})-\frac{F}{2\omega}\sin{(\Phi+\Delta t)} +\frac{\Gamma}{4\omega^2 A} \\
-\frac{F}{2\omega A}\cos{(\Phi+\Delta t)} 
\end{pmatrix}}_{\mathcal{F}(x)}+\underbrace{\begin{pmatrix}
1 & 0  \\
0 & A^{-1}
\end{pmatrix}}_{\mathcal{B}(x)}\Xi.\nonumber\\
{}
\end{eqnarray}}
\cc{To test} criterion \tr{(II)}, \cc{we verify that
\begin{eqnarray}
   \mathrm{skew}\Bigg( \underbrace{\begin{matrix}\begin{bmatrix}
\frac{\partial}{\partial A}   \\
\frac{\partial}{\partial \Phi}  
\end{bmatrix} \end{matrix}}_{\nabla}\,\,\underbrace{\begin{matrix} \begin{bmatrix}\begin{pmatrix}
1 & 0  \\
0 & A^{2}
\end{pmatrix} \mathcal{F}(x)\end{bmatrix}^T\end{matrix}}_{[\mathcal{B}^{-T}(x) \mathcal{B}^{-1}(x)\mathcal{F}(x)]^T}\Bigg)=\begin{pmatrix}
0 & 0   \\
0 & 0  
\end{pmatrix}.
\end{eqnarray}} 
\cc{Therefore,} Eqs. \eqref{example 1 eq 1}--\eqref{example 1 eq 2} have the form of a \sr{TLP} with $g(x)=\mathrm{diag}(1,A^2)$, $h(x)=\mathrm{diag}(1,A)$ and
\begin{eqnarray}
    \widetilde{\mathcal{V}}(x,t)=-\tr{\frac{A^2\lambda}{4}}+\frac{A^4}{32}+\frac{A F}{2\omega}\sin{(\Phi+\Delta t)}-\frac{\Gamma}{4\omega^2}\ln{A}. \label{vdp potential}\nonumber \\
    {}
\end{eqnarray}
\dc{The potential given by Eq. \eqref{vdp potential}\cc{, which was obtained by integrating $-\nabla \widetilde{\mathcal{V}}=\mathcal{B}^{-T} \mathcal{B}^{-1}\mathcal{F}$,} is visualized in Fig. \ref{Figure 2} for $t\in[0,2/\lambda]$, $\omega=5.03\times 10^3$, $\lambda$ equal to $2\%$ of $\omega$ and \lc{$\Gamma=\lambda^2\omega^2/10$} \lc{(arbitrary units)}. We observe a double-layered structure of the potential \lc{which is more pronounced at small values of $F/\omega \lambda^{3/2}$}, consistent with the notion that the \lc{(perturbed)} self-sustained oscillation coexists with the forced response at small forcing amplitudes.\cite[see][pp. 180--190]{balanov2009simple} We also note that, for nonzero detuning, the potential varies periodically in time, leading to beating oscillations, i.e., oscillations of the slow variables $A$ and $\varphi$.\cite[see][pp. 174--177]{balanov2009simple}}

\subsection{Generalized Kuramoto model \label{swarm section}}
\dc{A set of swarming oscillators (``swarmalators'') has been described by a \tr{generalized} Kuramoto model:\cite{okeefe}}
\begin{eqnarray}
    \dot{y}_i&=&\nu_i+\frac{\mathcal{\mathcal{J}}}{{\tr{N}}}\sum^{\tr{N}}_j\sin{(y_j-y_i)}\cos{(\theta_j-\theta_i)}, \label{y eq swarm}\\
    \dot{\theta}_i&=&\omega_i+\frac{\mathcal{K}}{{\tr{N}}}\sum^{\tr{N}}_j\cos{\tr{(y_j-y_i)}}\sin{(\theta_j-\theta_i)},\label{t eq swarm}
\end{eqnarray}
where $i=1,\dots,\tr{N}$, \tr{$n=2N$ is the system dimension,} \tr{the parameters} $\mathcal{J}$, $\mathcal{K}\in\mathbb{R}$ are coupling constants \tr{and} $\nu_i$, $\omega_i\in\mathbb{R}$ are the eigenfrequencies of the dynamic variables $y_i$ and $\theta_i$. To observe the exact potential, we define the new variables $x=(Y_1,\dots,Y_{\tr{N}},\Theta_1,\dots,\Theta_{\tr{N}})^T$, where $Y_i=y_i/\mathcal{J}$ and $\Theta_i=\theta_i/\mathcal{K}$. \dc{For $\mathcal{J}$, $\mathcal{K}>0$,} the resulting system is a \dc{TGS} with \tr{$g=\mathrm{diag}(\mathcal{J},\dots,\mathcal{J},\mathcal{K},\dots,\mathcal{K})\in\mathbb{R}^{\tr{n}}$} and
\tr{\begin{eqnarray}
    \widetilde{\mathcal{V}}(x)&=&-\sum^{N}_k\Big[\frac{1}{2N}\sum^{N}_{j}\cos{\mathcal{J}(Y_j-Y_k)}\cos{\mathcal{K}(\Theta_j-\Theta_k)} \nonumber\\
    &&+\nu_k  Y_k + \omega_k \Theta_k\Big]. \label{Full swarm potential}
\end{eqnarray}}
\dc{For $N=2$ and $\nu_i=\nu$ and $\omega_i=\omega$ for $i=1,2$, Eqs. \eqref{y eq swarm} and \eqref{t eq swarm} are equivalent to the following system describing the dynamics of the differences $\Delta y=y_2-y_1$ and $\Delta \theta=\theta_2-\theta_1$\cite[see][Eqs. (76) and (77)\lc{\footnote{\lc{Due to a typographical error, the minus signs are missing in Eqs. (76) and (77) of the reference.}}}]{okeefe}:
\begin{eqnarray}
    \Delta \dot{y}=-\mathcal{J}\sin\Delta y \cos \Delta \theta, \label{y diff eq.}\\
     \Delta \dot{\theta}=-\mathcal{K}\sin\Delta \theta \cos \Delta y. \label{theta diff eq.}
\end{eqnarray}
\begin{figure}[t!]
\begin{psfrags}
    
\psfrag{a}{}
\psfrag{b}{\hspace{0.43cm}$\log_{10}(\cc{\mathcal{K}})$}
\psfrag{c}{\hspace{-0.2cm}$-0.52$\hspace{1.61cm}$0$\hspace{1.61cm}$0.52$}
\psfrag{d}{\hspace{-0.55cm}$-0.52$\hspace{1.78cm}$0$\hspace{1.93cm}$0.52$}
\psfrag{e}{\hspace{0.98cm}$\log_{10}(\cc{\mathcal{J}})$}
\psfrag{1}{\hspace{-0.35cm}$\Delta \Theta$}
\psfrag{2}{\hspace{-0.1cm}$\Delta Y$}

    \centering
    \hspace{-0.05cm}\includegraphics[width=0.43\textwidth]{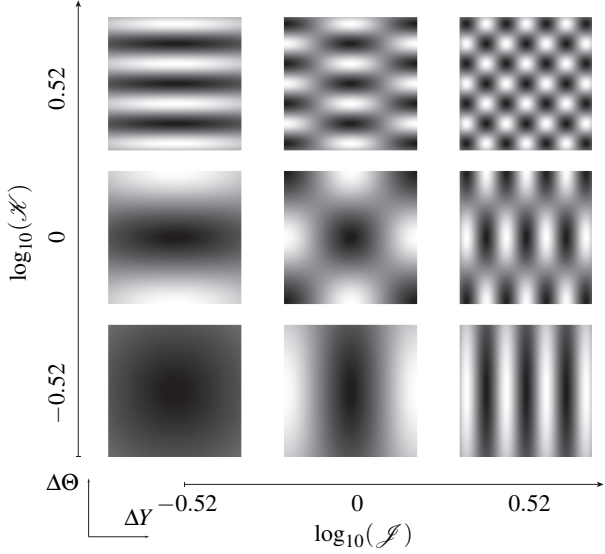}
\end{psfrags}
    \caption{\dc{Illustration of the results derived in Sec. \ref{swarm section}. Shown is the potential $\mathcal{V}$ given by Eq. \eqref{swarm potential} over the periodic domain $\Delta Y,\Delta\Theta\in[-\pi,\pi]$ for different values of the coupling constants $\mathcal{J}$ and $\mathcal{K}$ (logarithmic scale). The potential level is indicated in grayscale, ranging from black (attractive, $\mathcal{V}=-1$) to white (repelling, \cc{$\mathcal{V}=1$}).}}
    \label{Figure 3}
\end{figure}
Analogous to the general case above, we define the new variables $\Delta x=(\Delta Y,\Delta \Theta)^T$, $\Delta Y=Y_2-Y_1$ and $\Delta \Theta=\Theta_2-\Theta_1$ satisfying $\Delta \dot{x}=-g^{-1}\nabla \widetilde{\mathcal{V}}$ with
\cc{\begin{eqnarray}
    \widetilde{\mathcal{V}}(\cc{\Delta}x)=-\cos{(\mathcal{J}\Delta Y)}\cos{(\mathcal{K}\Delta\Theta)} \label{swarm potential}
\end{eqnarray}}
and the metric tensor $g=\mathrm{diag}(\mathcal{J},\mathcal{K})$.\cc{\footnote{\cc{Due to the preceding coordinate change to $\Delta x$, the potential given in Eq. \eqref{swarm potential} is not the same as that given in Eq. \eqref{Full swarm potential}, evaluated for $N=2$.}}} 

\lc{It is worth noting that the} system \eqref{y diff eq.} and \eqref{theta diff eq.} is known to possess an exact limit cycle.\cite[see][pp. 7--8]{okeefe} This solution is not in contradiction to the well-known theorem on nonexistence of periodic orbits in autonomous gradient systems,\cc{\cite[see][pp. 201--202]{strogatz2018nonlinear}} as it exists only for negative values of \lc{$\mathcal{K}$}, for which the metric tensor defined above loses its positive definiteness and the assumptions of our method break down.}

\dc{The potential \eqref{swarm potential} is visualized for different values of the coupling constants $\mathcal{J}$ \cc{and} $\mathcal{K}$ in Fig. \ref{Figure 3}. We see that the synchronized state $\Delta Y=\Delta \Theta=0$ is always a potential minimum and therefore linearly stable for $\mathcal{J}$, $\mathcal{K}>0$, which is consistent with an earlier stability analysis on the full system \eqref{y eq swarm} and \eqref{t eq swarm}.\cite[see][pp. 4--5]{okeefe}}

\subsection{Coupled limit cycles}
The weakly nonlinear amplitude-phase dynamics of two linearly coupled Van der Pol oscillators, derived using deterministic averaging, are given by\dc{\cite[see][Eq. (4.11)]{balanov2009simple}}
\begin{eqnarray}
    \dot{A}_1&=&\frac{\lambda_1}{2} A_1-\frac{1}{8}A_1^3+\frac{C}{2}(A_2\cos{\phi}-A_1),\label{coupled limit cycles eq 1}\\
    \dot{A}_2&=&\frac{\lambda_2}{2} \cc{A_2}-\frac{1}{8}A_2^3+\frac{C}{2}(A_1\cos{\phi}-A_2),\label{coupled limit cycles eq 2}\\
     \dot{\phi}&=&\Delta-\frac{C}{2}\sin{\phi}\Big(\frac{A_2}{A_1}+\frac{A_1}{A_2}\Big),\label{coupled limit cycles eq 3}
\end{eqnarray}
where $C$ is the coupling, $\phi=\varphi_2-\varphi_1$ is the phase difference and $\Delta=(\omega_{2}^2-\omega_1^2)/2\omega\approx\omega_{2}-\omega_1$ is the detuning between the eigenfrequencies $\omega_{1,2}$ and $\omega$ is the frequency of the synchronized coupled oscillators \cc{satisfying} $\omega\approx \omega_{1,2}$.
If we define $x=(A_1,A_2,\Phi)^T$, $\Phi=\phi-\Delta t$, then Eqs. \eqref{coupled limit cycles eq 1}--\eqref{coupled limit cycles eq 3} are equivalent \cc{to} a \dc{TGS} with $g^{-1}(x)=\mathrm{diag}(1,1,A_1^{-2}+A_2^{-2})$ and
\begin{eqnarray}
    \widetilde{\mathcal{V}}(x,t)&=&-\frac{\lambda_1 A_1^2+\lambda_2 A_2^2}{\tr{4}}+\frac{A_1^4+A_2^4}{32}\nonumber\\
    &&+\frac{C}{4}[A_1^2+A_2^2-2A_1A_2\cos{(\Phi+\Delta t)}]. \label{exact potential coupled limit cycles}
\end{eqnarray}
It is an open question whether the \tr{above} results can simplify the stability analysis \tr{of} Eqs. \eqref{coupled limit cycles eq 1}--\eqref{coupled limit cycles eq 3}.\cite[see][pp. \cc{80}--84 \cc{and pp. 409--415, respectively}]{Aronson1990403,balanov2009simple}
\subsection{Nonlinear coupling \label{nl coupling example}}
The following set of nonlinearly coupled \sr{amplitude and phase equations are obtained by averaging the equation describing the projection of turbulence-driven thermoacoustic dynamics onto two orthogonal} modes in an annular cavity\lc{\cite{Noiray2013}} (see \dc{Fig. \ref{Figure 4}}, left inset\cc{,} \tr{for a sketch} \dc{of such a cavity}):
\begin{eqnarray}
    \dot{A}&=&\nu A-\frac{3\kappa}{32}\big(3A^2+\tr{[2+\cos{(2\phi)}]}B^2\big)A \label{Noiray eq. 1} \nonumber \\
    &&+\frac{\Gamma}{4\omega_0^2 A}+\zeta_a,\\
        \dot{B}&=&\nu B-\frac{3\kappa}{32}\big(3B^2+\tr{[2+\cos{(2\phi)}]}A^2\big)B\nonumber \\
        &&+\frac{\Gamma}{4\omega_0^2 B}+\zeta_b, \label{Noiray eq. 2}\\
        \dot{\phi}&=&\frac{3\kappa (A^2+B^2)}{32}\sin{(2\phi)}+\Big(\frac{1}{A}+\frac{1}{B}\Big) \zeta_\phi, \label{phase diff eq 2 Noiray}
\end{eqnarray}

where \sr{$A$ and $B$ are the amplitudes, $\phi=\varphi_a-\varphi_b$ is the phase difference,} $\omega_0$ is the eigenfrequency, $\nu$ is the growth rate\sr{,} $\kappa$ is the nonlinearity constant \tr{\cc{and} $\Xi=(\zeta_a,\zeta_b,\zeta_\phi)^T$} \cc{contains} white noise sources of equal intensity $\Gamma/2\omega_0^2$. Equation \eqref{phase diff eq 2 Noiray} can be rewritten as two separate equations for the phases $\varphi_a$ and $\varphi_b$:
\begin{eqnarray}
     \dot{\varphi_a}&=&\frac{3\kappa B^2}{32}\sin{2(\varphi_a-\varphi_b)}+\frac{\xi_a}{A}, \label{phase eq 1 Noiray}\\
         \dot{\varphi_b}&=&-\frac{3\kappa A^2}{32}\sin{2(\varphi_a-\varphi_b)}+\frac{\xi_b}{B}, \label{phase eq 2 Noiray}
\end{eqnarray}
where $\xi_{a,b}$ are white noise sources of \cc{equal} intensity $\Gamma/2\omega_0^2$. Starting from Eqs. \eqref{phase eq 1 Noiray}--\eqref{phase eq 2 Noiray}, Eq. \eqref{phase diff eq 2 Noiray} can be derived by taking the difference between the \tr{former} two equations and setting $\xi_a=-\xi_b=\zeta_\phi$. 
According to criterion \tr{(II)}, Eqs. \eqref{Noiray eq. 1}, \eqref{Noiray eq. 2}, \eqref{phase eq 1 Noiray} and \eqref{phase eq 2 Noiray} correspond to a \sr{TLP} with $x=(A,B,\varphi_a,\varphi_b)^T$, $\Xi=(\zeta_a,\zeta_b,\xi_a,\xi_b)^T$, $g(x)=\mathrm{diag}(1,1,A^2,B^2)$, $h(x)=\mathrm{diag}(1,1,A,B)$ and
\begin{eqnarray}
    \widetilde{\mathcal{V}}(x)&=&-\frac{\nu(A^2+B^2)}{2}+\frac{3\kappa}{128}\big(3(A^4+B^4)\nonumber\\
    &&\tr{+2 A^2 B^2 [2+\cos{2(\varphi_a-\varphi_b)]}\big)}\nonumber\\
    &&-\frac{\Gamma}{4\omega_0^2}\ln{(AB)}.
\end{eqnarray}

\subsection{Quaternion flow \label{Abel example}}
\begin{figure}[t!]
\begin{psfrags}
    
\psfrag{a}{\hspace{0.005cm}\textcolor{arrowred}{$\theta$}}
\psfrag{b}{\hspace{-0.035cm}\textcolor{arrowgreen}{$2\chi$}}
\psfrag{c}{\hspace{-0.03cm}\textcolor{arrowblue}{$A$}}
\psfrag{d}{\hspace{-1.05cm}\textcolor{poleorange}{\normalsize{CCW spinning}}}
\psfrag{e}{\large{\hspace{-6.2cm}$\psi=\textcolor{arrowblue}{A}e^{i(\textcolor{red}{\theta}-\Theta)}e^{-k\textcolor{arrowgreen}{\chi}}e^{j(\omega_0 t +\varphi)}$}\hspace{0.65cm}\textcolor{poleblue}{\normalsize{CW spinning}}}
\psfrag{f}{\hspace{-0.3cm}\textcolor{planegray}{\normalsize{standing}}}
\psfrag{j}{\hspace{0.02cm}$\Theta$}
\psfrag{k}{}
\psfrag{l}{}
\psfrag{m}{(a)\hspace{3cm}(b)}

    \centering
    \includegraphics[width=0.45\textwidth]{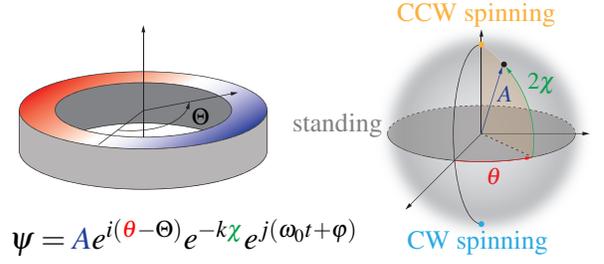}
\end{psfrags}
\caption{Illustration of the examples presented in \cc{Secs.} \ref{nl coupling example} and \ref{Abel example}. In \dc{this representation,} a self-oscillating \tr{mode} $\psi$ in an annular cavity is projected onto four variables $x=(A,\cc{\chi,\theta},\varphi)^T$ using the basic quaternions $(i,j,k)$ (left inset).\dc{\cite{Ghirardo2018,Faure-Beaulieu2020,Ghirardo2021,indlekofer_faure-beaulieu_dawson_noiray_2022}} Different states such as \tr{pure} spinning and standing waves are \dc{mapped to} different points on the Bloch sphere (right inset). The same coordinate system is used to represent the PDF isosurfaces in Fig. \ref{Figure 5}. }
    \label{Figure 4}
\end{figure}

\begin{figure}[t!]
\begin{psfrags}
    
\psfrag{a}{}
\psfrag{b}{\hspace{-0.45cm}noise intensity $\log_{10}\Pi$}
\psfrag{c}{\hspace{-0.4cm}$-1.77$\hspace{1.25cm}$-0.77$\hspace{1.5cm}$0.23$}
\psfrag{d}{\hspace{-0.22cm}$0$\hspace{1.88cm}$0.82$\hspace{1.88cm}$1.64$}
\psfrag{e}{\hspace{0.9cm}asymmetry $\gamma$}

    \centering
    \hspace{-0.05cm}\includegraphics[width=0.43\textwidth]{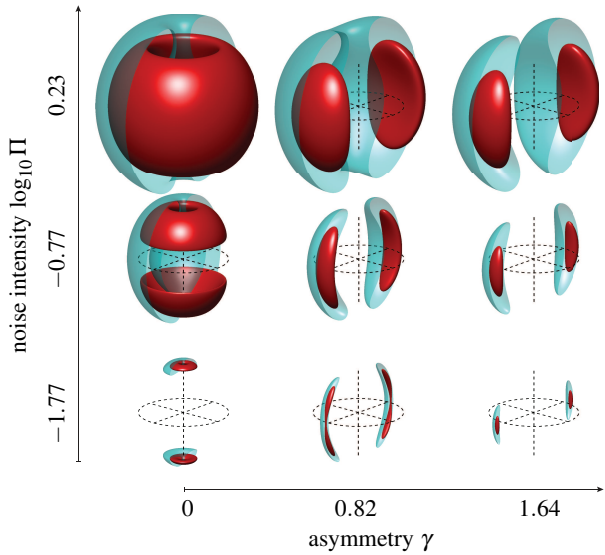}
\end{psfrags}
    \caption{\tr{Illustration of the results derived in Secs. \ref{Abel example}. Shown are} isosurfaces of the transformed \cc{steady-state} PDF $\widetilde{P}_\infty$ \tr{given by Eqs. \eqref{transformed steady sol} and \eqref{quaternion potential}} corresponding to \dc{$25\%$} (\cc{cyan, one half shown}) and \dc{$75\%$} (\cc{red}) of its maximum value, as a function of the nondimensionalized noise intensisity \tr{$\Pi$} and asymmetry $\gamma$ (semi-$\log$ scale, definitions in \dc{the main text}). The spherical coordinate system used to represent the PDF is defined in \dc{Fig. \ref{Figure 4}}. \tr{The length of the dashed vertical line and the dashed circle's \cc{diameter} are both equal to $\cc{16}\sqrt{\nu/15\kappa}$.} The spatial structure of the analytical Fokker--Planck solution shown in this figure is in excellent agreement with \dc{corresponding} numerical simulations.\dc{\cite[see][Fig. 11]{indlekofer_faure-beaulieu_dawson_noiray_2022}} }
    \label{Figure 5}
\end{figure}

In the study of self-oscillating \sr{thermoacoustic} \tr{modes} in annular cavities, \sr{an alternative projection to the one used in \dc{the previous example} and} based on \dc{the quaternion Fourier transform for bivariate signals\cite{Flamant2019351}} offers a \sr{convenient description of the nature of the modal dynamics, where one of the state variables indicates whether spinning or standing waves govern the dynamics at a given time instant}.\dc{\cite{Ghirardo2018,Faure-Beaulieu2020,Ghirardo2021,indlekofer_faure-beaulieu_dawson_noiray_2022}} \sr{Indeed, by} projecting the \sr{acoustic field} $\psi(\Theta,t)$ depending on the azimuthal angle $\Theta$ onto \sr{the} four \sr{state} variables $x=(A,\tr{\chi,\theta},\varphi)^T$ using the basic quaternions $(i,j,k)$, \sr{the instantaneous state can be mapped} to different points on the Bloch sphere.\dc{\cite{Ghirardo2018}} In this \sr{representation}, counter-clockwise (CCW) and clockwise (CW) spinning waves correspond to the north ($2\chi=\pi/2$) and south ($2\chi=-\pi/2$) poles, while the equatorial plane (\tr{$\chi=0$}) describes pure standing waves (\dc{Fig. \ref{Figure 4}}). In general, the system state is a mixture between a standing and a spinning wave. The variable $\theta$ describes the orientation of the nodal line of \lc{the standing wave component of} $\psi$. \dc{By deterministic and stochastic averaging of the projected acoustic wave equation, the following dynamics for $x$ can be derived:\cite[see][pp. 20--23]{Faure-Beaulieu2020}}
\begin{eqnarray}
    \dot{x}=\mathcal{F}(x)+\mathcal{B}(x)\Xi. \label{abel general sys}
\end{eqnarray}
The entries of the deterministic term $\mathcal{F}$ in Eq. \eqref{abel general sys} are
\begin{eqnarray}
    \mathcal{F}_1&=&\bigg(\nu+\frac{c}{4}\cos(2\theta)\cos(2\chi)\bigg)A\nonumber\\
    &&-\frac{3\kappa}{64}[5+\cos(4\chi)]A^3+\frac{3\Gamma}{4\omega_0^2 A}, \label{Abel eq. 1}\\
    \mathcal{F}_2&=&\frac{3\kappa}{64}A^2\sin(4\chi)-\frac{c}{4}\cos(2\theta)\sin(2\chi)\nonumber\\
    &&-\frac{\Gamma\tan(2\chi)}{2\omega_0^2A^2},\\
    \mathcal{F}_3&=&-\frac{c}{4}\frac{\sin(2\theta)}{\cos(2\chi)},\\
    \mathcal{F}_4&=&\frac{c}{4}\sin(2\theta)\tan(2\chi),\label{Abel eq. 4}
\end{eqnarray}
where $\nu$ is the growth rate, $\kappa$ is the nonlinearity constant, $\omega_0$ is the eigenfrequency and $c$ is the asymmetry. The entries of the noise vector $\Xi=(\zeta_A,\zeta_\chi,\zeta_\theta,\zeta_\varphi)^T$ each have equal intensity $\Gamma/2\omega_0^2$ and the matrix $\mathcal{B}$ describing the stochastic coupling is given in the reference as\lc{\footnote{\lc{Due to a typographical error, the matrix defined in Eq. (78) of the reference should instead be equal to its transpose.}}}
\begin{eqnarray}
       \mathcal{B}= \begin{pmatrix}
1 & 0 & 0 & 0\\
0 & A^{-1} & 0 & 0\\
0 & 0 & \dfrac{1}{A\cos2\chi} & 0\\
0 & 0 & -\dfrac{\tan2\chi}{A} & A^{-1}
\end{pmatrix}. \label{B matrix}
\end{eqnarray}
\dc{By criterion \tr{(II)}, Eqs. \eqref{abel general sys}--\eqref{B matrix} \tr{describe} a \sr{TLP} with}
\begin{eqnarray}
    \hspace{-0.2cm}{g}^{-1}(x)&=& \begin{pmatrix}
1 & 0 & 0 & 0\\
0 & A^{-2} & 0 & 0\\
0 & 0 & \dfrac{1}{A^2\cos(2\chi)^2}& \tr{-\dfrac{\tan2\chi}{A^2 \cos 2 \chi}}\\
0 & 0 & -\dfrac{\tan2\chi}{A^2 \cos 2 \chi} & A^{-2}+\dfrac{\tan(2\chi)^2}{A^2}
\end{pmatrix}, \nonumber \\
  \hspace{-0.2cm}{h}^{-1}(x)&=& \mathcal{B}(x), \nonumber
\end{eqnarray} 
and the transformed potential
\begin{eqnarray}
    \widetilde{\mathcal{V}}(x)&=&-\bigg(\nu+\frac{c}{4}\cos(2\theta)\cos(2\chi)\bigg)\frac{A^2}{2}\nonumber\\
    &&+\frac{3\kappa}{256}[5+\cos(4\chi)]A^4\nonumber\\
    &&-\frac{3\Gamma}{4\omega_0^2}\ln(A)-\frac{\Gamma}{4\omega_0^2}\ln{\big(\cos{(2\chi)}\big)}. \label{quaternion potential}
\end{eqnarray}
\tr{In \dc{Fig. \ref{Figure 5}}, we plot the transformed \cc{steady-state} PDF $\widetilde{P}_\infty(x)$ given by Eqs. \eqref{transformed steady sol} and \eqref{quaternion potential} in the \lc{spherical} coordinate system $(A,2\chi,\theta)$ on a semi-$\log$ scale} for different values of the nondimensionalized noise intensity \tr{$\Pi$} and asymmetry $\gamma$\tr{\,:}
\begin{eqnarray}
    \tr{\Pi}&=&\frac{27\kappa\Gamma}{256\nu^2\omega_0^2}, \label{PI}\\
    \gamma&=&\frac{c}{2\nu}. \label{GAMMA}
\end{eqnarray}
As $\gamma$ is increased from zero, a preferred direction in $\theta$ emerges in the steady state, demonstrating the explicitly broken symmetry of the system for $\gamma\neq 0$. The spatial structure of the analytical PDF $\widetilde{P}_\infty$ shown in \dc{Fig. \ref{Figure 5}} is in excellent agreement with numerical simulations of the Fokker--Planck equation \dc{for the same parameter values}.\dc{\cite[see][Fig. 11]{indlekofer_faure-beaulieu_dawson_noiray_2022}}
\section{Conclusions}
In this study, we derived necessary and sufficient criteria for the existence of an exact potential in a general noise-driven system. We demonstrated on several broadly studied \fc{models} of deterministic and stochastic \fc{oscillations} that \dc{from the differential-geometric properties of} transformed \sr{potential systems driven by additive \cc{white} noise}\dc{, one can derive} new analytical descriptions of their nonlinear dynamics. \cc{The potentials and PDFs obtained in this work may be used in the future to investigate the corresponding \cc{models} from a different perspective, for example by visualizing families of trajectories with different initial conditions and relating their nonlinear dynamics to the potential landscape.} 

\dc{Systems to which \cc{the} method \cc{presented in this work} applies appear to be ubiquitous \cc{in the literature} and are often found in the context of time-averaged flows.} The question \dc{of} \textit{why} \dc{this is the case} is left for future research to answer. \dc{To conclude, we \cc{also} mention that our theoretical approach implies a self-consistent way of modeling noise in given deterministic gradient flows, and that the resulting models are exactly solvable if the potential is stationary.}
 
\section*{Author's contributions}
\cc{\textbf{Tiemo Pedergnana}: Formal analysis (lead), visualization (lead), writing--original draft, writing--review and editing (equal). \textbf{Nicolas Noiray}: Formal analysis (supporting), supervision, visualization (supporting), writing--review and editing (equal).}

\section*{Acknowledgements}
\dc{\lc{The authors} acknowledge helpful discussions with Kevin O'Keeffe about his previous work on swarming oscillators. This project is funded by the Swiss National Science Foundation under Grant agreement 184617.}

\section*{AUTHOR DECLARATIONS}
\subsection*{Conflict of Interest}
\dc{The authors have no conflicts to disclose.}
\section*{Data availability}
\dc{The datasets used for generating
the plots in this study can be directly obtained
by numerical simulation of the related mathematical equations in the manuscript.}

\bibliography{aipsamp}

\end{document}